\documentclass{jaa}
\usepackage{natbib}
\usepackage{graphicx}
\usepackage{url}

\newcommand{\A}{\emph{AstroSat}}
\newcommand{\N}{N_{\rm{points}}}
\newcommand{\M}{M_{\rm{sum}}}
\newcommand{\Np}{N_{\rm{pairs}}}
\newcommand{\ps}{\rm{\, s^{-1}}}

\begin{document}\sloppy

%%paper title
%%For line breaks \\ can be used within title
\title{Characterisation of Cosmic Ray Induced Noise Events in \A -CZT Imager}

%%author names are separated by comma (,)
%%use \and before the last author name
%%use a * along with the number separated by comma
%% for the  author for correspondence
%%\textsuperscript{number} is used for affiliation
%%\affilOne, \affilTwo etc., upto \affilTwentyfive is possible
%%Please note the first letter after \affil is capitalised in the command
%%

\author{Debdutta Paul\textsuperscript{1,2},  A. R. Rao\textsuperscript{1,3,*}, Ajay Ratheesh\textsuperscript{1,4,5},   N.P.S. Mithun\textsuperscript{6}, S.V. Vadawale\textsuperscript{6}, Ajay Vibhute\textsuperscript{3}, Dipankar Bhattacharya\textsuperscript{3}, Priya Pradeep\textsuperscript{7},  and S. Sreekumar\textsuperscript{7}}
%\author{AUTHOR1\textsuperscript{1}, AUTHOR2\textsuperscript{1} and AUTHOR3\textsuperscript{2,*}}
\affilOne{\textsuperscript{1}Tata Institute of Fundamental Research,
            Homi Bhabha Road, Colaba, Mumbai, 400005, India.\\}          
\affilTwo{\textsuperscript{2}Gubbi Labs LLP, No: 2-182, 2nd Cross, Extension, Gubbi, Karnataka, 572216, India.\\}
\affilThree{\textsuperscript{3}Inter University Centre for Astronomy \& Astrophysics, Post Bag 4, Ganeshkhind, Pune,  411007, India.\\}
\affilFour{\textsuperscript{4} Department of Physics, Tor Vergata University of Rome, Via della Ricerca Scientifica 1, I-00133 Rome, Italy\\}
\affilFive{\textsuperscript{5} INAF - IAPS, Via Fosso del Cavaliere 100, I-00133 Rome, Italy.\\}
\affilSix{\textsuperscript{6}Physical Research Laboratory, Navrangpura, Ahmedabad, 380009, India.\\}
\affilSeven{\textsuperscript{7}Vikram Sarabhai Space Centre, Kochuveli, Thiruvananthapuram,  695022, India.\\}

%%escape two column mode for title, affiliation and abstract
%%by giving \twocolumn command as shown

\twocolumn[{

\maketitle

%%include \corres to print the corresponding author Email id
\corres{a.raghu.rao@gmail.com}

%%include \msinfo for
%%manuscript information such as
%%received, revised and accepted dates
%%
\msinfo{ }{ }

%%abstract
\begin{abstract}
The Cadmium Zinc Telluride (CZT) Imager onboard \A\, consists of pixelated CZT detectors, which are sensitive to hard X-rays above 20 keV. The individual pixels are  triggered  by ionising events occurring in  them, and the detectors operate in a self-triggered mode, recording each  event  separately   with information about its time of incidence, detector co-ordinates, and channel that scales with the amount of ionisation. The detectors are  sensitive not only to photons from astrophysical sources of interest, but also prone to a number of other events like background X-rays, cosmic rays, and noise in detectors or the electronics. 
In this work a detailed analysis of the effect of cosmic rays on the detectors is made and it is found that cosmic rays can trigger multiple events which are  closely packed in time (called `bunches'). Higher energy cosmic rays, however, can also generate delayed emissions, a signature previously seen in the PICsIT detector on-board INTEGRAL.  An algorithm to automatically detect them based on their spatial clustering properties  is presented.
%Some of the events are identified as due to   high energy cosmic rays, 
%which produce signatures previously seen in the PICsIT detector on-board INTEGRAL.   
Residual noise events are examined   using examples of Gamma Ray Bursts as target sources.

\end{abstract}

%%insert keywords separated by 3 hyphens using \keywords{words}

}]
%%close the twocolumn escape here

%%include \doinum{number}for the DOI number in the header
%%include \volnum{number} for the volume number in the header
%%include \year{yyyy} for  year of publication in the header
%%include \pgrange{num--num} page range of article in the header
%%include \artcitid{num} for the article citation id
%%include \lp to print last page of the article
%%include \setcounter{page}{pagenum} for the exact starting page of the article

\doinum{12.3456/s78910-011-012-3}
\artcitid{\#\#\#\#}
\volnum{000}
\year{0000}
\pgrange{1--}
\setcounter{page}{1}
\lp{1}

\section{Introduction}
\label{sec:Intro}

\A\, is a broad band high energy Indian mission covering UV, soft X-rays and hard X-rays \citep{Singh_etal.2014, Rao_et_al.-2016-arXiv-Astrosat}. It comprises four co-aligned  instruments to cover a wide bandwidth: Ultra Violet Imaging Telescope/UVIT \citep{Subramaniam_et_al.-2016-SPIE, Tandon_et_al.-2017-AJ, Tandon_et_al.-2017-JApA, Rahna_et_al.-2017-MNRAS}, Soft X-ray Telescope/SXT \citep{Singh_et_al.-2016-SPIE, Singh_et_al.-2017-JApA}, Large Area X-ray Proportional Counter/LAXPC \citep{Agrawal_et_al.-2017-JApA, Antia_et_al.-2017-ApJS, Yadav_et_al.-2017-arXiv} and Cadmium Zinc Telluride (CZT) Imager/CZTI \citep{Bhalerao_et_al.-2017-JApA}. CZTI is a hard X-ray instrument sensitive in the energy range $20$-$200$ keV, consisting of an array of CZT detectors. Each detector module consists of 256 independent detectors, called pixels, of nominal size $2.5$ mm $\times \; 2.5$ mm and 5 mm thickness. The CZT plane consists of four quadrants, each with 16 detector modules, comprising an effective area of $ 1024 \, \rm{cm{^2}} $. CZTI has imaging capabilities below $100$ keV, using the Coded Aperture Mask (CAM) placed above collimator slats that surround the detector modules. The collimator is made of Tantalum of size $4$ cm $\times \; 4$ cm, which allows for a field of view (FoV) of $ 4.6^\circ \times 4.6^\circ $. In addition to spectroscopic, timing, and localization capabilities \citep{Rao_et_al.-2016-ApJ}, CZTI can measure  polarization in the hard X-rays with exposure time an order of magnitude smaller than previously existing instruments \citep{Chattopadhyay_et_al.-2017-arXiv}. Further, CZTI acts as an all sky monitor above $\sim$ 100 keV enabling it to do spectro-polarimetric studies of gamma-ray bursts \citep{Rao_et_al.-2016-ApJ},  making it a unique instrument at these energies.

The details of CZTI including overall instrument configuration, the detectors and electronics, the data characteristics, processing pipeline and default products have been given  in  \cite{Bhalerao_et_al.-2017-JApA}.
In CZTI terminology, an `event' is a trigger in any of the pixels, characterized by a unique time-stamp, the pixel co-ordinates, and the `pulse height amplitude' or PHA, which is linearly related to the amount of ionisation in the triggered  pixel. 
The cause of ionisation is mainly due to X-ray photons from the targeted X-ray sources along with the  background X-rays (cosmic X-ray background as well as locally generated by cosmic rays).   In addition to this, 
ionisation is also caused by interactions of charged particles in the detector.
As  CZTI is a hard X-ray instrument, for most sources including the Crab Nebula, the number of background events are much more than the source events. These background events are mostly  induced by the charged particle environment of the detector, which, other than showing predictable variation with the satellite location, should show   a uniform random  distribution in spatial and temporal dimensions. 
%Additionally, there are  events caused by  the d
Direct interaction of charge particles in the detector can cause a heavy ionisation  resulting in multiple events  and  
events may also be generated by  peculiarities in the detector. These additional events (other than those caused by the interaction of X-ray photons) are called `noise' events and they  need to be identified and removed before doing any scientific investigation of the events from the target source, both in the temporal and energy domains.

The presence of cosmic ray induced noise events was deduced during the first few days of the mission, and an algorithm is implemented in the   ground analysis software (hereafter called the CZTI pipeline) to eliminate them from the science data. They are easily distinguished from science events due to their temporal characteristics: they all `bunch' together within the smallest interval of time resolvable by the CZT detector electronics, $ 20 \rm{\, \mu s} $ \citep{Bhalerao_et_al.-2017-JApA}. A steady stream of these `bunches',   defined as three or more events with a time separation between successive events no more than one time stamp  ($ 20 \rm{\, \mu s} $), are observed in the data and they are understood as due to   cosmic rays  continuously bombarding the detector plane.   The  number of such bunches that should occur from background X-rays by chance is $\sim  10^{-3}$ bunches per second whereas about 70 bunches per second are seen in the data.  These bunches temporally track the variation of the cosmic rays bombarding the entire satellite, independently measured by the Charged Particle Monitor (CPM) on-board \A\, \citep{Rao_et_al.-2017-JApA}.
It was also known, before the launch of the satellite, that some of the pixels could be prone to electronic noises. The CZTI pipeline identifies these pixels, removes the events from these pixels from further analysis, keeping track of the effects on live time and effective area. If any pixel shows a consistently  noisy behaviour, they are permanently disabled by a ground command.  The subtle effects of a pixel becoming electronically noisy immediately after the bombardment of a high energy charge particle in that  pixel is also taken care of in the current CZTI pipeline.  Hence the pipeline removes both types of noise viz., charge particle induced and pixel misbehaviour: the former by removing the bunches and the post-bunch noise events and the latter by identifying the misbehaving pixels in a given observation and completely eliminating  data from these pixels.

In this work, we present the results of an investigation of the noise events in CZTI. In particular, we have examined whether all cosmic ray induced events could be identified by temporal bunching alone. It was found that
the   heavy deposition of charge in the detector modules by very high energetic particles can be  identified via patterns on the detector modules that are characteristic of pixelated detectors collecting data at such high time resolutions. A new algorithm is developed which automatically identifies and removes these events from the data. 
The manifestation  of cosmic rays as spatial structures in the data  is  investigated in the next section,
 via an algorithm detailed in the Appendix. 
 In Section 3, we discuss these spatial structures as manifestations of higher energy cosmic rays. 
 In  the last section,  concluding remarks are given along with some test results.

 \section{Examination of spatial structures in the data  }
 \label{sec:Structures}

  For the analysis presented here, we
 use the astronomer-friendly `Level 2' (L2) FITS files created by the Payload Operation Centre (POC) of CZTI, located in the Inter-University Centre for Astronomy \& Astrophysics (IUCAA) in Pune, India.  The analysis employs the standard procedure outlined in the CZTI  pipeline User Guide available at the AstroSat Science Support  Cell (\url{http://astrosat-ssc.iucaa.in}).  However, we  have also used the `Level 1' FITS files to repeat the exercise for some of the data sets.
The details of the data structure and   this ground data analysis pipeline can be found in the companion paper (Ajay Ratheesh et al., this volume). 

\subsection{Data preparation}

The CZTI pipeline uses several parameters to clean the data. The major thrust of cleaning  the cosmic ray induced noise events in the data is based on the   observation  that long lingering noise events can occur in certain pixels after the incidence of  cosmic rays,  particularly for the higher energy cosmic rays. Further, events from  certain pixels  that show abnormal behaviour throughout the observations (gross noisy pixels) or at specific times (flickering pixels) are  typically discarded for the length of the observation or for specific durations, respectively.  
For the present work, we make a conservative use of the event removal  so that we can attempt a spatial identification of the residual noise.  We therefore remove  only the events from the gross noisy pixels and retain the rest of the data.

\begin{figure*}
\begin{centering}
\includegraphics[scale=0.5]{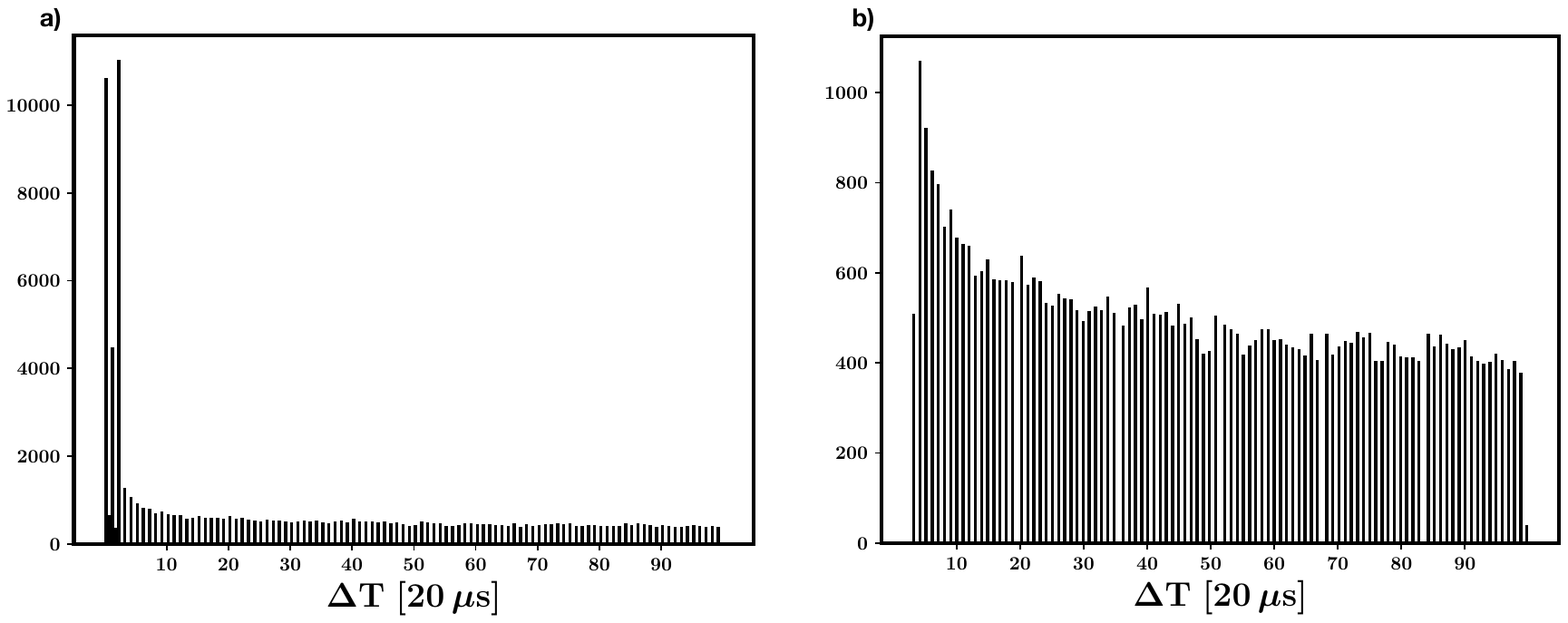}
\end{centering}
\caption{A histogram of the number of bunches within time $\Delta T$ of a preceding bunch.    \textbf{a)}:   The histogram peaks at small $\Delta T$  showing that some bunches are connected to the subsequent bunches.  The bunches are redefined as a single bunch if the time separation is less than $t_2$.  \textbf{b)}: With $ t_2 = 60 {\rm \,\mu s} $,  this initial peak is removed.
% Although the above plots are taken from one orbit of March background data, this observation is true for all datasets examined. The first point in corresponds to the bunch redefinition timescale, and is an artefact created due to the limitation of the way division is carried out in the binary system; it persists whatever value of $t_2$ is used, but is unimportant for our purposes.
\label{fig:bunch_redefinition}}
\end{figure*}

The understanding behind the idea of  these temporal bunches is that each bunch is created by one cosmic ray particle generating a series of electronic events within timescales shorter than the instrumental resolution of $ 20 {\rm \,\mu s} $.  It is to be noted that the   $ 20 {\rm \,\mu s} $ time bin is  digitally generated but the onboard  analogue electronics is capable of recording events at a much faster time scale ($\sim$ $ 1 {\rm \,\mu s} $) and then serially transmit the events. If such is the case, then  the time of occurrence of   respective bunches is  independent of each other, and the interval between one bunch and the next, $\Delta T$, is expected to follow  a flat temporal distribution. When plotting the histogram of $\Delta T$  using the bunch data, it is clearly seen that such is not the case (Fig. \ref{fig:bunch_redefinition} a). This  observation leads one to assume that the  effect of cosmic rays sometimes can last longer than the instrumental time resolution of 20 $\mu$s, used for defining bunches. 
Hence, we  redefine bunches such that, if the interval between one bunch and the next is less than a certain duration $t_2$, then these two bunches are understood to be created by the same cosmic ray particle,   and  all data of these two bunches as well as data within the two are considered to belong to a single bunch.
Empirically, it is seen that the sharp spike  observed in Figure 1a is removed by  choosing $ t_2 = 60 {\rm \,\mu s} $  (see  Figure 1b).
It is observed that about  $10 \%$ of the  bunches are  added together by this method.

\begin{figure*}
\begin{centering}
\includegraphics[scale=0.5]{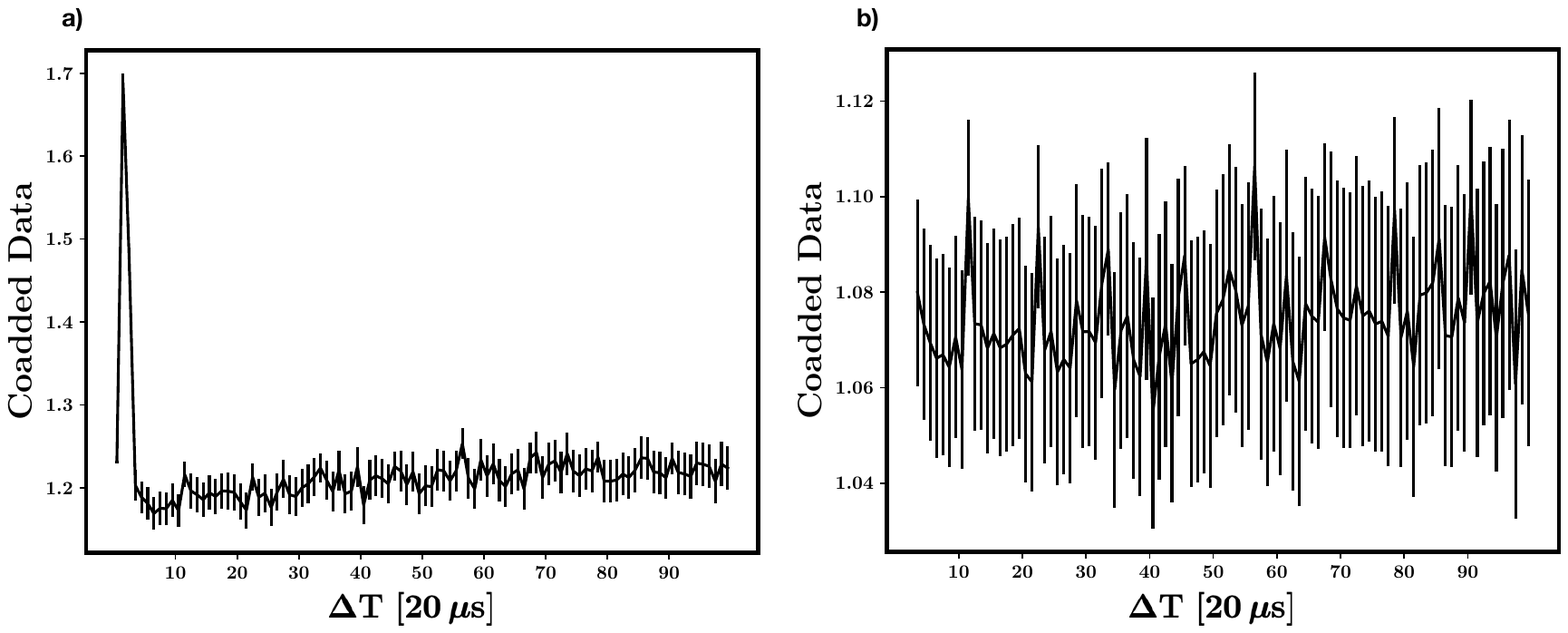}
\end{centering}
\caption{ {\bf a)}: The co-added light curve of events immediately after a bunch, normalised to the exposure time. A sharp spike  is seen at timescales less than $ 60 {\rm \,\mu s} $. 
% if co-added light curve, corrected for the exposure, is made from data post bunches, even after redefining bunches. This implies that post bunches, significant amount of electronic noise, created by the bunches, is persistent. Hence, When data post bunches are removed up to the parameter $t_3$. 
{\bf b)}: After removing  all data events after time $ t_3 (= 60 {\rm \,\mu s})$, post bunches,  the co-added light curve looks flat, all the way up to $2$ ms. All errors assume that the data are  Poissonian. %which is strictly true for the part away from the spike in \emph{Left}.
\label{fig:postbunch_flagging}}
\end{figure*}

To remove   the  effects of cosmic-ray particles that persist for some amount of time post-bunch after initially triggering a series of events in the detector, we also discard  events up to time    $t_3$ after the end of a bunch. %, similar to \emph{skipT1}, \emph{skipT2}, \emph{skipT3}. To estimate the optimum value of $t_3$, 
In Figure 2 we have plotted the co-added light curves of events after the bunches.
  That is, we generate a lightcurve of events from the end of a bunch to the beginning of the next bunch. We then repeat this for all the bunches, and redefine the start time for each lightcurve as the end of the first bunch. Then, we add all these data, which we refer to as co-adding.
Although each bunch lightcurve can be of different lengths, the process of co-adding takes care of the different durations. This is because the resultant lightcurve for each bunch is renormalised by the amount of live time available for the bin.
 In Figure \ref{fig:postbunch_flagging} we show this re-normalised post bunch light curve.
If the data is fully Poissonian, then it should vary randomly about unity, for all times. In   Figure 2a, we see a sharp rise 
and fall at the smallest time interval, which is indicative of remnant  effects even after co-adding the bunches.
  On choosing  $ t_3 = 60 {\rm \,\mu s} $ and removing data for $t_3$ after each bunch, the co-added light curve  becomes flat around unity. Hence, all gross features of the cosmic ray induced events can be removed by assuming that the post bunch effect lasts for $t_3$, although subtle after-effects of bunches at longer time scales cannot be completely  ruled out.

\begin{figure}
\begin{center}
\includegraphics[scale=0.5]{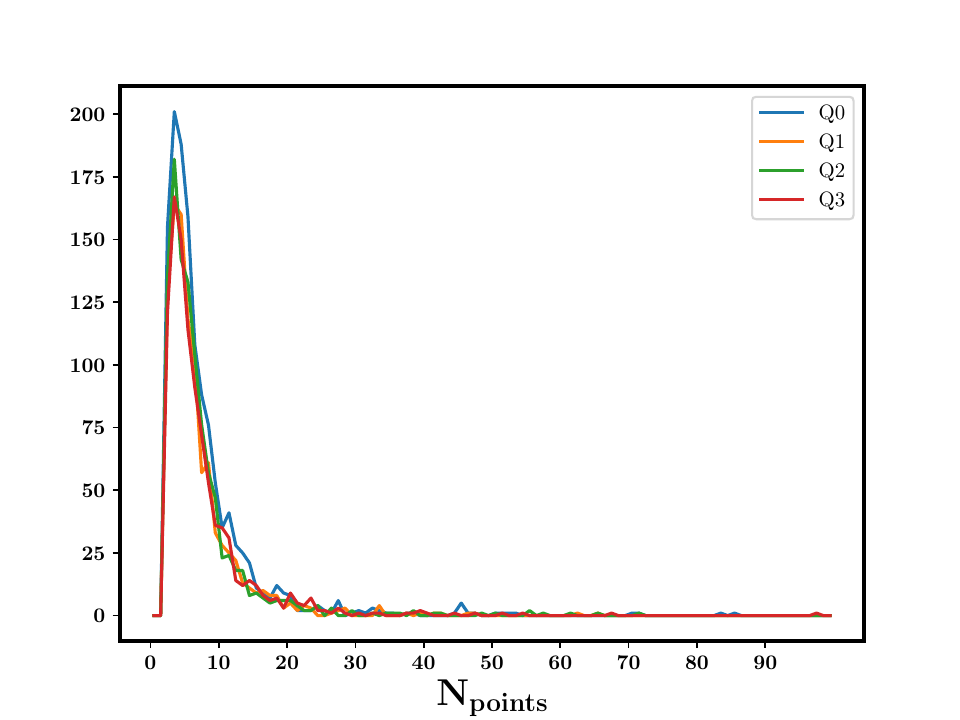}
\end{center}
\caption{A histogram of the number of noise structures, called DPHstructures, obtained from the algorithm DPHclean, as a function of the number of unique pixels  in them ($\N$), for an   observation using CZTI  lasting for an orbit, given separately for each quadrants (Q0 to Q3).
\label{fig:histogram_of_observed_Npoints}}
\end{figure}

\subsection{DPHclean}
\label{sec:DPHclean}

In the light curves made from the data after the procedure described above, %bunchclean and removing gross noisy pixels, 
strong temporary features, on a time scale of  hundred milliseconds, are observed.
 To investigate this further,  we examine the 
 detector  plane histograms (DPHs) of the  events  that create these features. 
DPH, for a given length of time,  is created by adding together all counts registered in a pixel and
showing all pixels in a geometric plane, with appropriate colour coding for the number of events. 
DPH created for a full observation is used to identify noisy pixels and a series of DPH, created every 100 $\mu$s, 
shown as a movie
(such a movie can be seen in the CZTI website \footnote{\url{astrosat.iucaa.in/czti/images/CZTI_Crab_5s.mpg}}, which highlights the impact of cosmic rays in CZTI data. The DPH 
created for 100 ms, when a spike in the light curve is observed (see for example Figure A4 in the Appendix),
shows that the excess  events cluster in some parts of the detector plane rather selectively. The timescale for detecting such clustering is examined, parametrized by $t_{look}$, the bin size for the light curve generation. Initially, such  spatial clustering is observed to be present for $5$-$\sigma$ outliers in light curves binned at $100$ ms. The events that contribute to the clustering are spread over timescales less than $100$ ms, and only very rarely involve two consecutive bins of $100$ ms.  We have developed  an algorithm 
called `DPHclean'  for the automatic identification of such  spatial clustering in the DPHs, henceforth called `DPHstructures'.   This algorithm is specifically suited for CZTI data where a) the disturbance due to cosmic ray interaction is a few pixels wide, b)  a limited number of pixels are affected by the interaction c) the geometric regions where the disturbance occur could be a few and d) the time scale of disturbance  could be several tens of milliseconds. We should note here that while the particle interaction time scale is in nanoseconds, the electronic data reading time scale is in micro-seconds, and such fast disturbances are already seen as bunches. As discussed earlier,   the   $ 20 {\rm \,\mu s} $ time bin is  digitally generated but the onboard  analogue electronics has a much faster time scale ($\sim$ $ 1 {\rm \,\mu s} $). The examination of the data at longer time scales is motivated by the existence of such structures in the data.    The details are given in the Appendix.  Since this algorithm is independent of the total number of events in the DPH, the only constraint on $t_{look}$ (optimised to $100$ ms) is  that it should be more than the duration of such events. 
 We also note here that the DPHclean specifically caters to the need
of the simultaneous two-event data indicative of Compton scattering by keeping aside all events satisfying the Compton
criteria: simultaneous double events from neighbouring pixels.

A large fraction of the events clustered in DPHstructures occupy regions only a few pixels wide. A histogram of DPHstructures with respect to the number of pixels   or points ($\N$) in them (for an   observation lasting for an orbit) is shown in  Fig. \ref{fig:histogram_of_observed_Npoints}. Some of these DPHstructures 
include pixels which register counts $4$ or higher in $100$ ms bins. Note that in case a DPH shows clustering, only those events 
in the DPH that are responsible for the clusters are removed from the data by DPHclean. 
The algorithm thus has the ability to identify the clusters, and 
%in the presence of such clustered events 
even during bright GRBs 
 algorithm selectively picks out only the events in the cluster and removes them, without 
affecting the GRB photons. The advantage of such selective identification is evident. It is noticed that running this algorithm on \emph{all} DPHs made from a given set of data reduces the noise significantly more than selectively running it on (say $5$-$\sigma$) outliers in the light curve. 
%In  Fig. \ref{fig:light curve_comparison_of_bunches_and_DPHstructures} are plotted the light curves of bunches and DPHstructures, both binned at $100$ s intervals for a full orbit data, given separately for the four quadrants of CZTI (Q0 to Q3).  
We have examined the light curves of bunches and DPHstructures and we see an  overall rising trend, in both the light curves, when   the satellite enters the South Atlantic Anomaly (SAA). % is clear for bunches, whereas for DPHstructures, it is only marginal. However, t
 This similarity points to the origin of both kinds of events in charged particles. The quadrant-averaged orbit-averaged rate of DPHstructures is $0.25 \ps$.
%It is hypothesized that they are locally generated electronic noise or tails of cosmic ray bunches lingering in the data for timescales longer than bunches. On the other hand, it is most likely that the DPHstructures that cause strong outliers in light curves are caused by a physical mechanism that gives rise to genuine ionization in the detectors, and not random electronic noise.

\begin{figure}
\begin{center}
\includegraphics[scale=0.25]{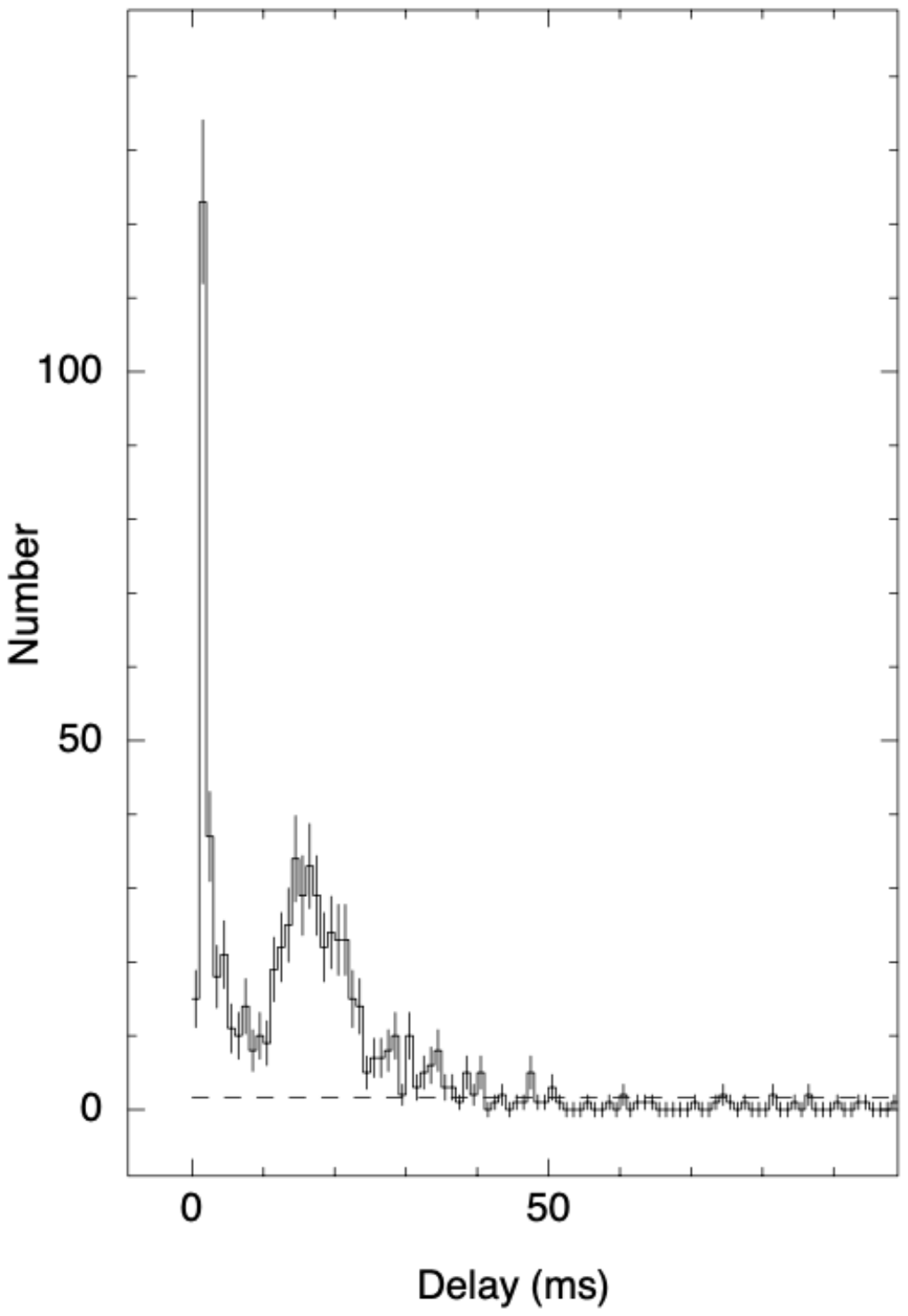}
\end{center}
\caption{ Histogram of the number of  DPHstructures  as a function of the time difference between the start of a super bunch just prior to the DPHstructures and the start of the DPHstructure. The horizontal dashed line is the expected number of chance coincidences.
%{\emph {Top}}: when the time difference between the heavy bunch and DPHstructure is less than 2 ms and {\emph {Bottom}}: when the time difference is between 2 and 30 ms.
 \label{fig:HeavyDelay}}
\end{figure}

\begin{figure}
\begin{center}
\includegraphics[scale=0.3]{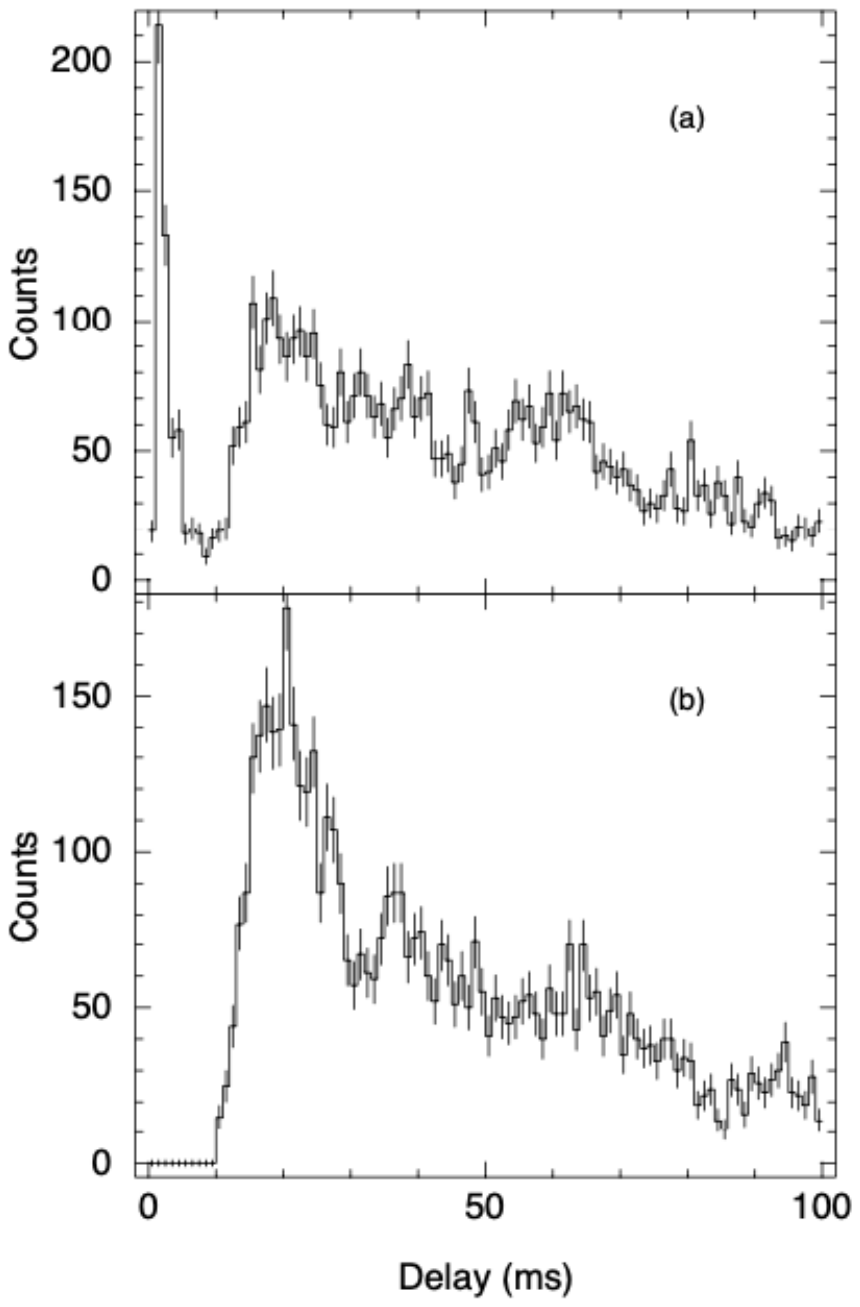}
\end{center}
\caption{ Light curve of   the  events in a DPHstructure  as a function of the time difference between the start of a super bunch just prior to these DPHstructures, {\bf (a)}:  for DPHstructures with a delay from the preceding super bunch less than 5 ms, and {\bf (b)}: for DPHstructures with a delay from the preceding super bunch  between 5 ms and 25 ms.
\label{fig:Delay}}
\end{figure}

We have investigated any possible temporal correlation of bunches with DPHstructures, to understand whether DPHstructures could be caused by  bunches.  The DPHstructure rate is $0.25 \ps$ and their duration is in tens of milliseconds, whereas the bunches are more numerous ($70 \ps$) and are of very short duration ($\sim$0.1 ms), hence, there will be a large number of chance associations in millisecond timescales. Therefore, we have examined the occurrence of the start of a DPHstructure after a `super bunch', 
 defined as a bunch 
 with the participating number of events more than 100 (see Ajay Ratheesh et al., this volume). The rate of such super bunches is about  $1.5 \ps$. 
 In Figure 4, we show a histogram of the start time of    DPHstructures plotted with respect to the end time of an immediately preceding super bunch,
 for DPHstructures observed during one orbit of data for one quadrant. The expected numbers purely by chance are shown as a dashed line in the figure. The association is extremely significant: the probability of so many DPHstructures clustering within 25 ms of the super bunches, purely by chance, is negligibly small. We note here that a similar distribution is obtained for the other quadrants too and the distribution of super bunches occurring after a DHPstructure is as expected by chance.
 
It can be seen that there are two types of associations. Firstly, we find that a substantial number  ($\sim$ 20\%) of DPHstructures are closely associated with the super bunches: they occur within 5 ms of a super bunch (see Figure 4),  indicating that these DPHstructures could be post bunch noises caused by super bunches. 
 % The expected rate by chance  is 0.15\% and the chance probability for our sample is less than 10$^{-13}$. 
 %Hence, we can  conclude that these  DPHstructures are nothing but the long duration 
%post bunch noises caused by heavy bunches. 
Secondly, there is a peak between 5 and 25 ms in the delay histogram (for about 30\% of the DPHstructures - see Figure 4) along with a very significant dip   at around 10 ms. 
 
We have examined the lightcurves of the events participating in the DPHstrucures. We make a lightcurve for each DPHstructure, renormalise the start time with respect to the end time of a super bunch just preceding the concerned DPHstructure and co-add all the light curves together. Such lightcurves are made separately for the DPHstructures which have a delay from a super bunch   less than 5 ms (the first type discussed above -- Figure 5a) and for DPHstructures  having a delay between 5 ms and 25 ms (the second type -- Figure 5b). For the first type we see a sharp fall in the events initially and then a slow profile. The initial surge of events  perhaps indicates the post bunch noise. It is interesting to note the clear morphological differences in the temporal structure of the events  between these two types of associations. For DPHstructures with a delay less than 5 ms from a super bunch, 
there is a surge of events initially showing that the concerned DPHstructures could be post bunch noise in the affected pixels. On the other hand, for the second type, there is an interesting trend in the events participating in the DPHstructures. They have a typical time scale of 50 ms to 100 ms and they start generating  events 10 ms after the occurrence of the super bunch.

Since we identify each bunch as due to cosmic rays and since each cosmic ray is expected to be independent of other cosmic rays, about 50\% of DPHstructures are causally connected to super bunches. Some of them are post bunch noises while the majority of them are delayed emissions. The remaining 50\% of the DPHstructures could be a combination of these two: post bunch noises  from other bunches or separate emissions with a slow rise morphology. A cursory glance of the co-added light curves of these DPHstructures indicate that they may be dominated by the delayed emission morphology.
 
%For the purpose of the pipeline flow, however, it is safe to remove all DPHstructure events.

\section{DPHstructures as high energy cosmic ray events}
\label{sec:DPHstructures}

\begin{figure*}
\begin{centering}
\includegraphics[scale=0.6]{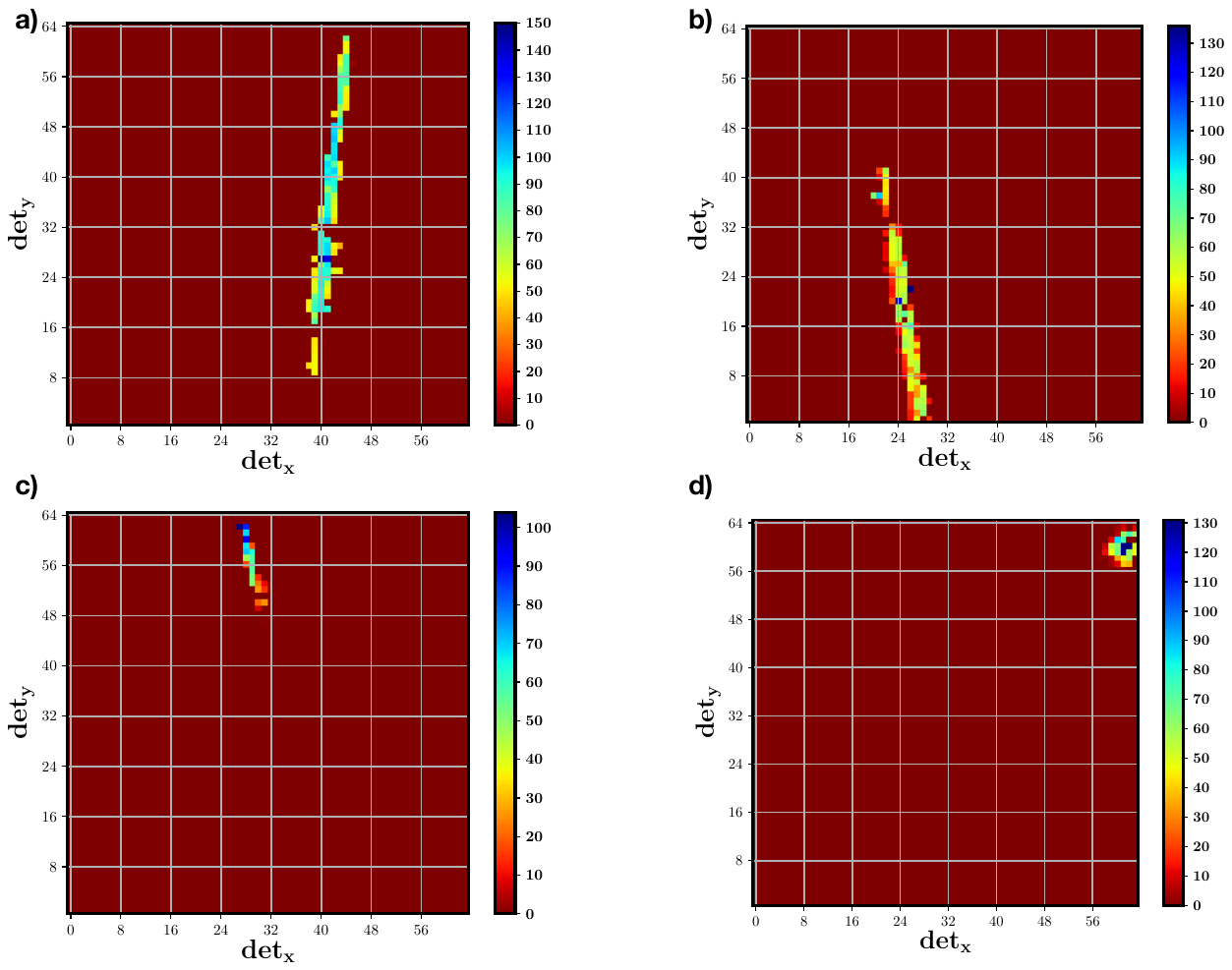}
\end{centering}
\caption{Detector Delay Histograms: Plotted in colour are the delay of the particular event from the first event in the cluster, in milliseconds, as a function of the position in the detector plane. These examples last unusually long, covering two consecutive time bins (100 ms). In the top two panels  (a and b), we observe linear tracks, the delay increasing along the track in \emph{(b)}. The delay patterns in   the bottom two panels   (c and d) - are remarkably similar to those seen in PICsIT on INTEGRAL due to phosphorescence-decays from events triggered by cosmic ray showers.
\label{fig:DDH}}
\end{figure*}

 \cite{Segreto_et_al.-2003-A&A--INTEGRAL_cosmic_rays} studied the detector characteristics of the PICsIT detector plane of the IBIS instrument on board INTEGRAL, and found events similar to DPHstructures. To investigate their cause, they plotted detector delay histograms (DDHs) corresponding to each DPHstructure. DDHs are histograms, on the detector plane, of the delay of the events contributing to a particular DPHstructure, with respect to the first event. They found evidence of two kinds of events: linear tracks, and a particular kind of delay pattern-- a gradual increase of the delay towards the centre of the  elliptical patterns seen in the DDHs. They explained these by the bombardment of the detector plane by charged particles or cosmic ray showers generated by hadronic and leptonic processes very close to the detector. They demonstrated that the delay in the first and last events in a particular DPHstructure being $ \sim 100 $ ms  could be explained by the saturation of the pixels by the extreme high energies of the charged particles, the pattern on the detector tracing the density of the cosmic ray showers in the logarithmic scale. Inspired by these findings, we plotted the DDHs for our DPHstructures, some examples of which are given in Fig. \ref{fig:DDH}. We see two kinds of events:

\begin{enumerate}

\item Those tracing linear tracks indicating trajectories of physical entities along them (Fig. \ref{fig:DDH}, a,b ). This points to the origin being charged particles which deposit their energy over multiple pixels that fall on their trajectory of motion through the detector.

\item Those with the delay being more in the inside of a cluster compared to its  spatial boundaries (Fig. \ref{fig:DDH}, \emph{c, d}).

\end{enumerate}

Both kinds of events are in striking similarity with the DDHs observed in PICsIT on board INTEGRAL, and naturally leads one to the hypothesis that DPHstructures in CZTI are also created by the bombardment of high-energy charged particles or cosmic ray showers.

\begin{figure}
\begin{center}
\includegraphics[scale=0.5]{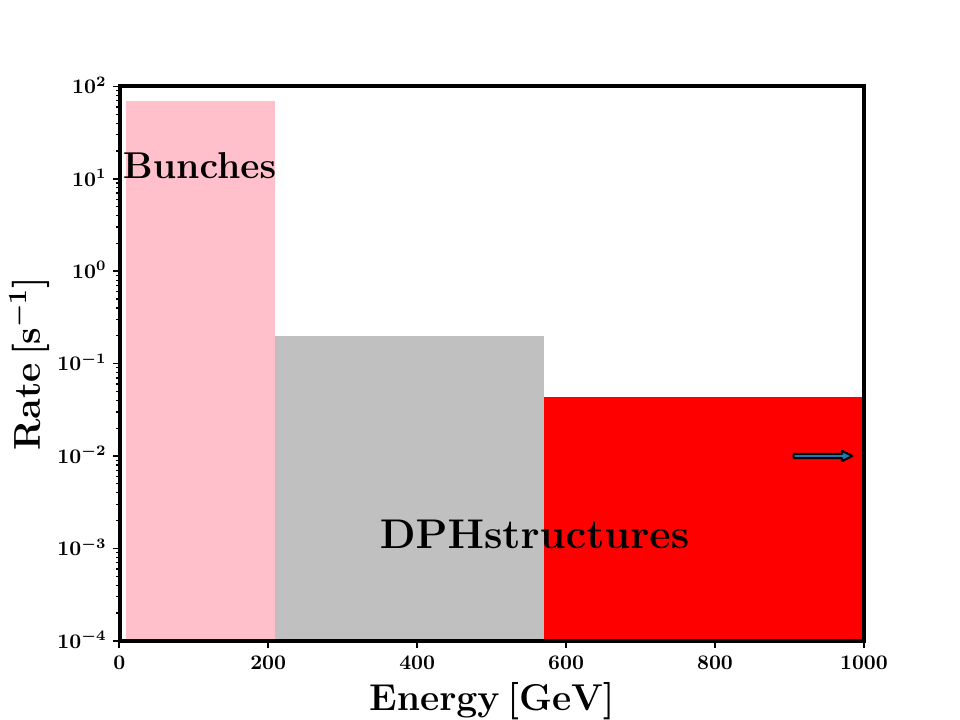}
\end{center}
\caption{The limits of the energies of three kinds of events: bunches, high-frequency DPHstructures, and low-frequency DPHstructures,  shown in pink, grey and red colours, respectively. The frequency cut of DPHstructures is based on the number of unique points in the DPHstructures, and is put roughly at the start of the tail of this curve, see Fig. \ref{fig:histogram_of_observed_Npoints}. The higher end of the low-frequency cosmic rays is used as $100$ TeV, but shown here only until $1$ TeV for representational purposes.
The arrow figuratively shows this extension of the energy limit.
\label{fig:Energy_limits}}
\end{figure}

Based on the similarity of the DDH found in CZTI, for DPHstructures,  with those in PICsIT onboard the INTEGRAL satellite, we can conclude that some cosmic rays, being more energetic than their counterparts that create bunches, might deposit their energies over multiple pixels instead of a few, thus resulting in DPHstructures. %, they are above the detectable energy threshold of the pixels, thus saturating them. Some cosmic rays might deposit their energies over multiple pixels instead of a few because being more energetic than their counterparts that create bunches,
 They appear to saturate these pixels, and, when the pixel output current drops below a threshold,  the affected pixels start registering events. The saturation timescale observed in the detectors then corresponds to the delay from the onset of the events created by these cosmic rays on the detector plane. The fact that DPHstructures are much less frequent than bunches also corroborates such a hypothesis. Further, it naturally explains the delay pattern of the second kind, i.e., those with progressively higher delays towards the centre of the  spatial pattern \citep{Segreto_et_al.-2003-A&A--INTEGRAL_cosmic_rays}. When a cosmic ray shower hits the detector, the density of the particles in the shower are traced by the delay in the DDH. The delay timescales in the detector pixels are thus deduced to be a few $ 100 $ milliseconds.

We can thus formulate the following scenario for the effect of the interaction of cosmic rays in CZTI:

\begin{enumerate}

\item  Low energy cosmic rays interact in a few pixels and the heavy charge deposition is cleared out by the fast electronics in a series of events called bunches. 

\item  When the charge deposition is high, it has a paralysing effect on the concerned pixels, and these pixels give out events in a slow time scale, manifested as DPHstructures.

\item In some cases, multiple interaction is also possible, giving rise to multiple centres in the DPHstructure as well as a DPHstructure existing together with a bunch.  We emphasise again that Compton scattering giving simultaneous  interactions is treated separately as Compton events and does not affect the bunch and DPHstructure analysis presented here.

\item Additionally, some bunches can also generate post-bunch electronic noise events.

\end{enumerate}

The energy of the cosmic rays cannot be calculated directly. However, we place constraints on the energy of both bunches and DPHstructures from the observed rate of the events, assuming a standard spectrum of cosmic rays \citep{Longair-3rd_Ed.}:

\begin{equation}
\frac{dN}{dE} =  1.8 \times 10^4 \rm{\dfrac{nucleons}{s \, m^2 \, sr \, GeV}} \left( \frac{E}{1 \, \rm{GeV}} \right)^{-2.7}.
\label{eq:cosmic_ray_spectrum}
\end{equation}

%We assume that the DPHstructures illuminate $10$ pixels on an average, which is an area of $160  \,\rm{cm^2}$, and
By taking the total area of the detector and  integrating over all solid angles, between energy limits $E_{\rm{min}}$ and $E_{\rm{max}}$, we  match the resultant counts   with the observed rates of bunches and DPHstructures to determine  $E_{\rm{min}}$ and $E_{\rm{max}}$. For the purpose of continuity, we divide the DPHstructures into two kinds of events on the basis of their frequency, with the criterion being the number of unique points in the DPHstructure $\lessgtr 10$.  The orbit-averaged, quadrant averaged, rate of bunches is  $70 \ps$. Similar rates for the  high-frequency and low-frequency DPHstructures are, respectively,  $0.200 \ps$, and $0.044 \ps$.  Assuming an upper limit of the low-frequency DPHstructures as $100$ TeV, the energy-limits thus derived are shown in Fig. \ref{fig:Energy_limits}.

What is the physical mechanism that creates the $ \sim 100 $ ms timescale saturation effect in the pixels? For PICsIT, the timescale was explained by fluorescence states of the CsI detectors, which is not possible for the CZT detectors in CZTI. 
%The only explanation is the following: 
In CZTI, the high voltage (HV) to the detectors is provided through an RC network (with a time constant of  $ \sim 100 $ ms). 
The extreme high energy of the cosmic rays that hit the individual detectors can momentarily  reduce the HV such that the charges are not collected and the events are not registered. The time required for the HV to stabilise and for further events to be registered is  $ \sim 100 $ ms.  
%ets current pass through the RC-circuit that provides stability to the source of power to these detectors. That is, due to the extremely high energy deposited in the individual detectors in an extremely small time, they successfully exchange power from the power source, thus remaining saturated until the impending RC-circuit has stabilized. Then the timescale of the saturation is given by the time-constant of the impending RC-circuit, which is $ \sim 100 $ ms 
%(\textbf{ADD REFERENCE}). Unfortunately, there is no way to directly corroborate such a hypothesis, and we are forced to leave it there.

\section{Discussion and conclusions}
\label{sec:conclusions}

\begin{figure*}
\begin{centering}
\includegraphics[scale=0.7]{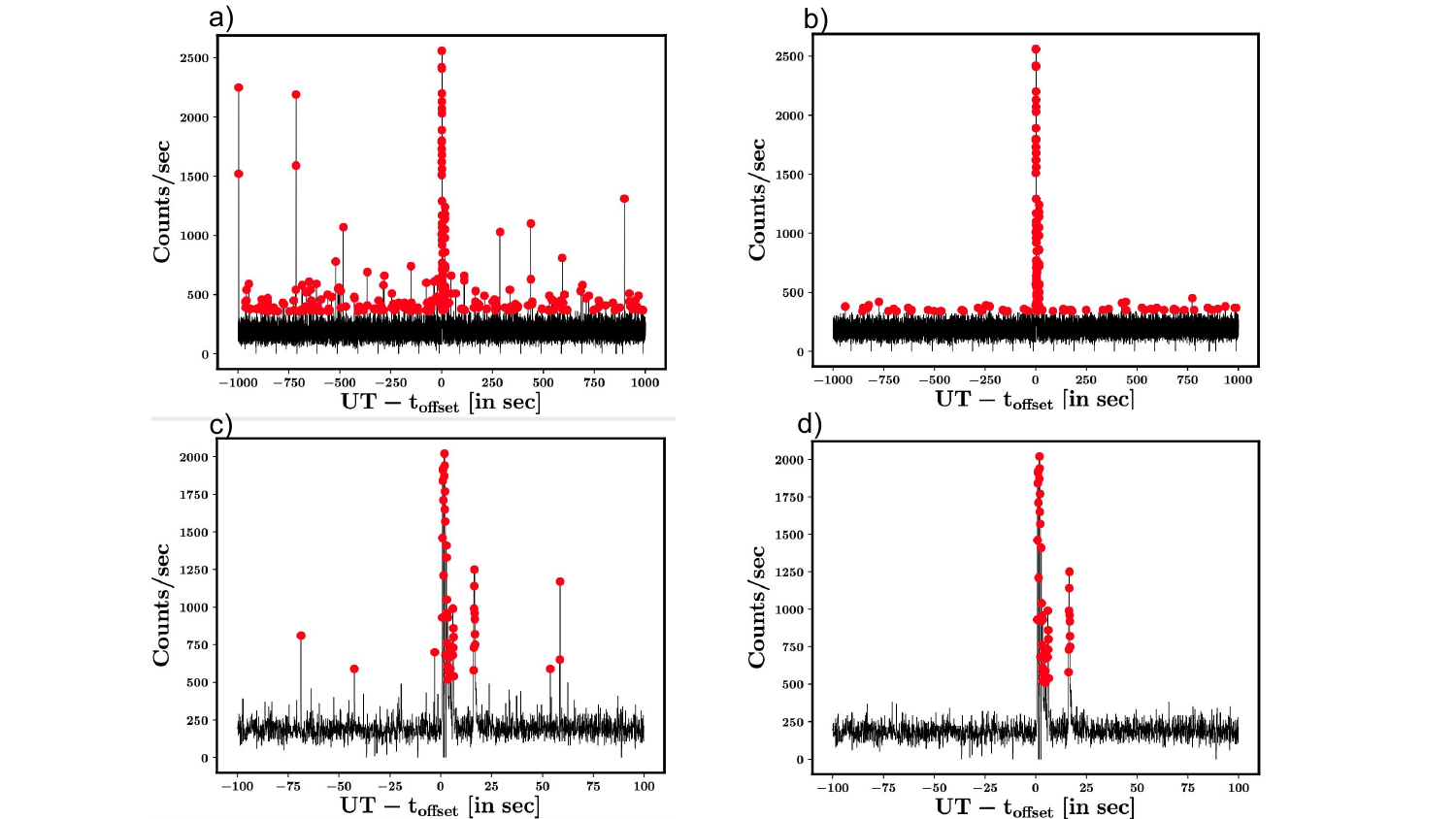}
\end{centering}
\caption{Lightcurves binned at $t_{\rm{look}} = 100$ ms before  (\emph{Left} - a and c) and after (\emph{Right} b and d) DPHclean near the bright GRB160802A, red points showing $2$-$\sigma$ outliers. In the top two panels (a and b)  longer stretches  of data from Q1 are shown whereas the bottom two panels (c and d) show the zoomed data around the GRB from Q0 (hence scale is different). 
%It is noted that sudden features in the light curve are removed by DPHclean, but GRB photons are not flagged. The second feature within $20$ seconds of the start of the prompt emission is also a part of the GRB, and is seen distinctly in the zoomed light curves of all quadrants with a similar profile, so it is not be mistaken as noise.
\label{fig:GRB_zoom}}
\end{figure*}

The long time scale (up to 100 ms) of the disturbance from  some cosmic rays is quite difficult to handle in the conventional dead time method: ignoring all data for these intervals would have increased the dead   time considerably. The method described here, DPHclean, however, removes the affected events selectively. Since only a small number of pixels are affected, the effective area reduction is extremely negligible. 
We show here the efficacy of DPHclean by showing a light curve near a GRB in Fig. \ref{fig:GRB_zoom}. 
Lightcurves binned at   100 ms (t$_{\rm{look}}$)  centred around the time $t_{\rm{offset}}$, corresponding to the trigger time of bright GRB160802A, are shown along with $2$-$\sigma$ outliers highlighted in red. A longer stretch of data from Quadrant 1 (Q1) is shown in the top panel (Fig 8 a,b) and a zoomed picture (for Q0, with a different scale), is shown in the bottom panel  (Fig 8 c,d).  The left panels (Fig 8 a,c) are the light curves before applying DPHclean and the right panels  (Fig 8 b,d) are after discarding all events responsible for DPHstructures. 
It can be seen  that noisy  features in the light curve are removed by DPHclean, but GRB photons are not flagged. The second feature within $20$ seconds of the start of the prompt emission is also a part of the GRB, and is seen distinctly in the zoomed light curves of all quadrants with a similar profile, and  it should  not be mistaken as due to noise.

In conclusion, we have understood the various effects of cosmic rays on the CZTI data. The new method, DPHclean, effectively removes the events from the affected pixels and this algorithm can be easily included in the ground analysis software. A new event cleaning method based on these and other considerations has been developed and it is described in detail elsewhere (Ajay Ratheesh et al.,  this volume).

\appendix
\setcounter{figure}{0} \renewcommand{\thefigure}{A.\arabic{figure}}
\section{Algorithm for detecting `DPHstructures'}
\label{Appendix}

\begin{figure*}
\begin{centering}
\includegraphics[scale=0.5]{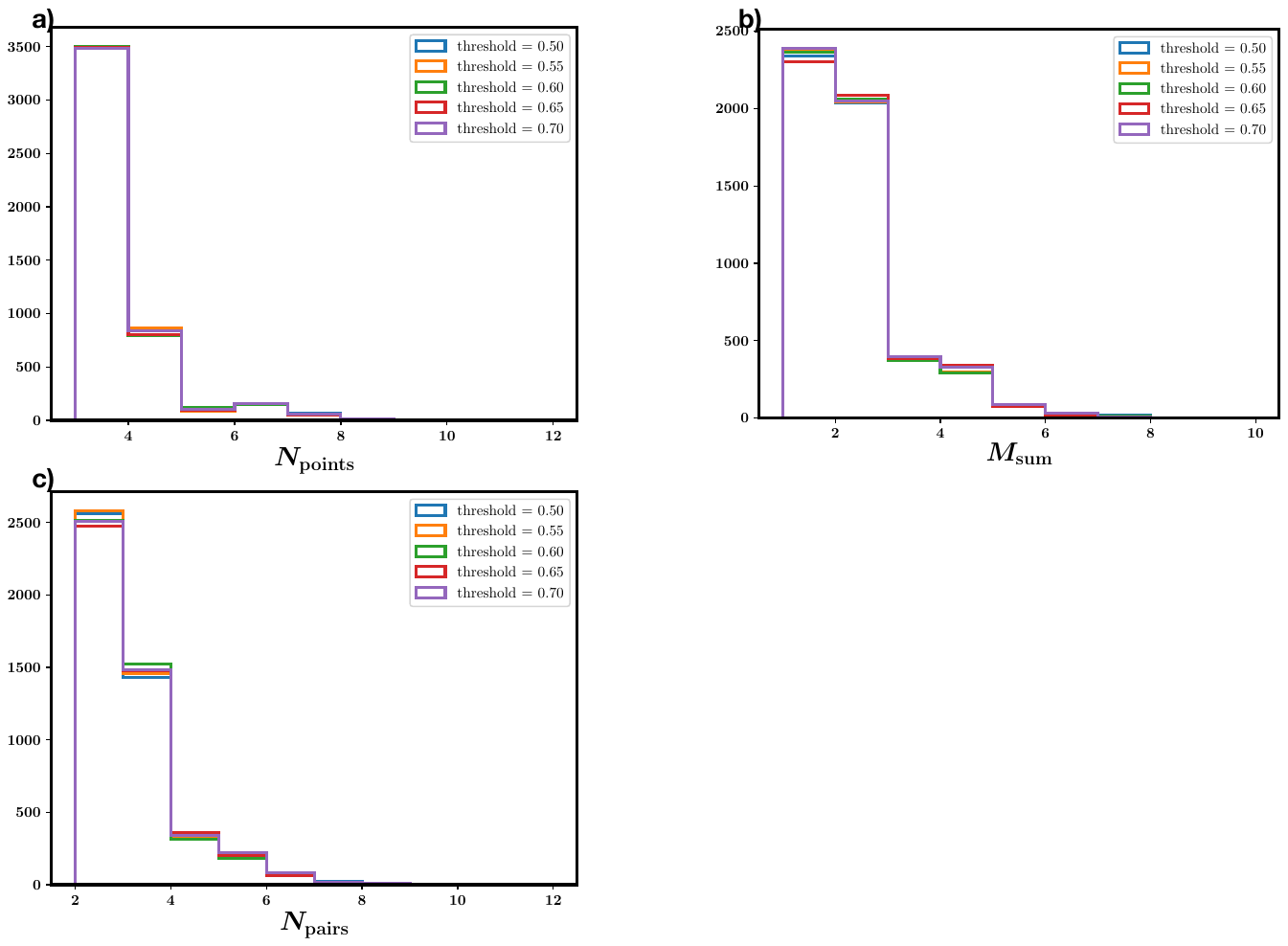}
\end{centering}
\caption{Histograms of {\bf (a)}  $\N$, {\bf (b)}  $\M$, and {\bf (c)}  $\Np$ with different values of $threshold$. Since most of the pairs are non-identical, $\N$ is more likely to be even than odd, hence it shows regular dips at odd integers  (as can be seen as slight dips in (a)). It is observed that these parameters are insensitive to the value of $threshold$ chosen. Hence, the  $threshold$ is fixed it at its most conservative upper limit throughout the work.
\label{fig:threshold_insensitiveness}}
\end{figure*}

\begin{figure*}
\begin{centering}
\includegraphics[scale=0.5]{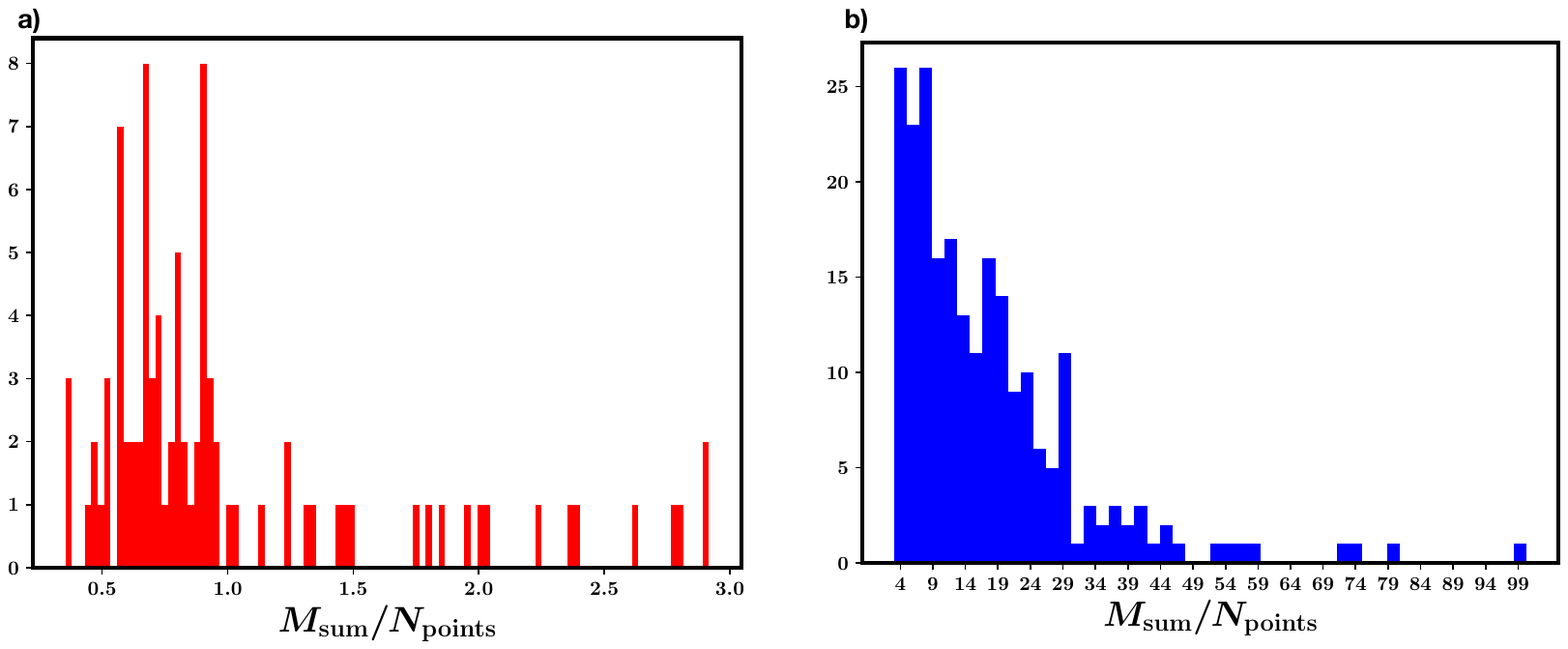}
\end{centering}
\caption{{\bf (a)}: Histogram of the parameter ${\M}/{\N}$ for the DPHs which are \emph{not} flagged.  This ratio rarely goes close to $3$, which is the value of $allowable$ used for flagging. {\bf (b):} Same for the ones that \emph{are} flagged. Note that although the total number of flagged DPHs in a typical dataset is much smaller than the number of DPHs that are flagged, most of them do not have any hot pairs, hence both $\M$ and $\N$ are zero. {\bf Fig (a)}  includes only those within finite values of $\M$ and $\N$, explaining why the total number is smaller than in {\bf Fig (b)}.  The sharp increase at values close to $4$ in {\bf (a)}, compared to the rarity of those in {\bf (b)} below $3$, demonstrates that the distinction is real. The reality of this distinction is also verified by manual examination of each DPH for long stretches of data which preferentially include weak as well as bright GRBs, examples of which are given in Fig. \ref{fig:DPHs_flagged_and_unflagged}.
\label{fig:robustness_of_allowable=3}}
\end{figure*}

\begin{figure}
\begin{center}
\includegraphics[scale=0.5]{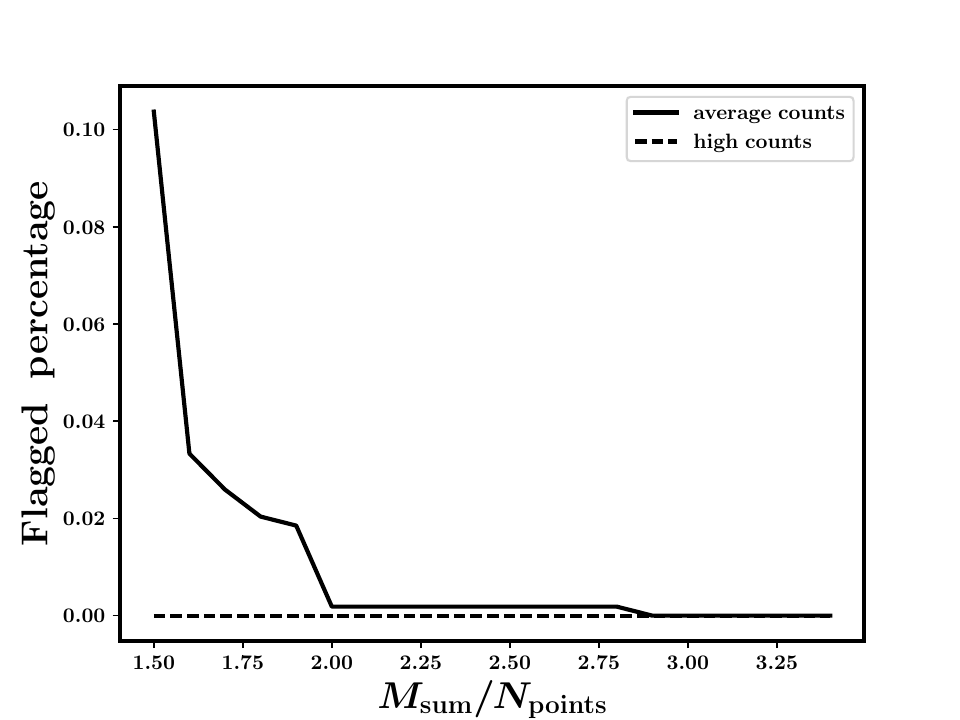}
\end{center}
\caption{Random DPHs are simulated with the event-rate as input. The parameter $allowable$ is allowed to vary, and flagging is carried out on the random DPHs based on these variable values. The plot show the resulting number of DPHs flagged as a percentage of the total number of DPHs simulated ($5400$), as a function of the variable.  An average count-rate of $150$ per second, and {\bf a}  high count-rate during bright GRBs, of $1500$ per second, are considered. The flagged percentage is $0$ for $ \M / \N > 3 $, proving the robustness of the algorithm for both the average data and during GRBs. In fact, it is more robust when the  count-rates are high as seen during GRBs, i.e. DPHs during bright GRBs have $\sim 0$ probability of getting flagged.
\label{fig:robustness_with_countrate}}
\end{figure}

\begin{figure}
\begin{centering}
\includegraphics[scale=0.6]{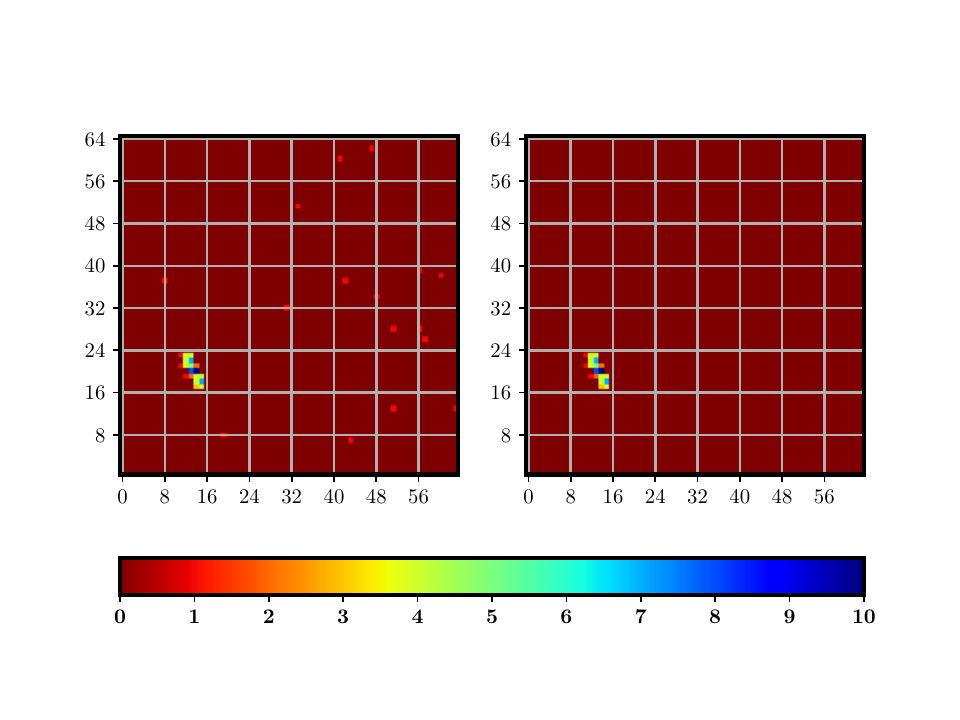}
\end{centering}
\begin{centering}
\includegraphics[scale=0.6]{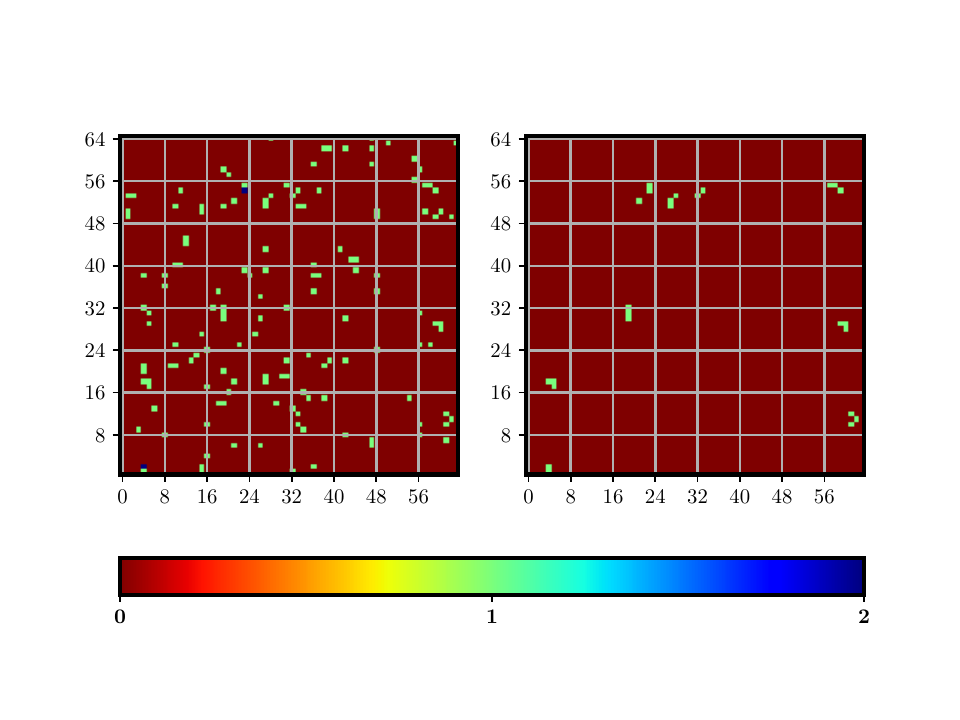}
\end{centering}
\caption{ Instructive examples of Detector Plane Histograms (DPH), plots of counts shown at positions detx vs dety.  \emph{Top}: On the \emph{left} is a DPH which shows clustering (the color-bar is of counts), with the identified clustered events shown in the \emph{right}. \emph{Bottom}: On the \emph{left} is a DPH that does not show clustering. The pairs that are used to test the clustering are explicitly shown in the \emph{right} to demonstrate that the presence of such random pairs are not enough to flag this DPH.
\label{fig:DPHs_flagged_and_unflagged}}
\end{figure}

The aim of such an algorithm is to consider a DPH, and numerically decide whether the DPH shows any  spatial  clustering or not. This algorithm will output a flag $0$ if clustering is detected or $1$ if it is not. A requirement of such an algorithm is that it should be independent of the total number of events in the DPH, since it is to be run on DPHs made during average count-rates as well as during GRBs.  The basic assumption made here is that genuine X-ray photons (from the astrophysical sources of interest as well as background X-ray photons extraneous to the instrument) are independent events and are randomly distributed in the detector plane, whereas  the `noise' events, ether induced by cosmic rays or from local electronic effects, would be  spatially clustered in the detector plane.

The algorithm is detailed below:
\begin{itemize}
\item Consider only those pixels in the DPH which register non-zero counts. If there are $n$ such pixels, there are $^{n}C_{2}$ pairs. For each pair, calculate a measure of `hotness', \[ m{}_{ij} = \dfrac{c{}_{i}\times c{}_{j}}{D_{ij}}, \] where $c_{i}$ is the count in the $i^{th}$ pixel and $D_{ij}$ is the distance between the pixels (in units of detx/dety, which is unity). This quantity is large if count in either pixel is large and/or if the distance between the pixels constituting the pair is small.

\item If \[ m{}_{ij} > threshold, \] call it a `hot pair'. It is to be noted that the maximum allowable value for threshold is 
\[ threshold_{\mathrm{max}} = \dfrac{1\times1}{\sqrt{2}} \simeq 0.707, \] which is the case for two diagonally-located neighbouring pixels registering $1$ count each. If $threshold$ is larger than this, we will miss these hot pairs, thus defeating the purpose.

\item Construct the set of all pixels which contribute to any such hot pair.

\item Modify the choice of hot pairs: if a pair is such that it consists of two neighbouring pixels only, each registering one count, and there is no hot pixel in its immediate neighbourhood, then do not consider the pair for the following steps. This ensures that actual double events are not considered whereas neighbouring pixels with one count each in the neighbourhood of a cluster are retained.

\item Calculate the `gross' parameters: 
\begin{enumerate}
\item total number of \emph{non-identical} points contributing to the identified hot pairs: $\N$;
\item the sum of the measures of the hotness for each such hot pairs: 
\[ \M = \sum_{\{\mathrm{all\, pairs}\}}m_{ij}\,; \]
\item the number of hot pairs detected (note that even if one pixel contributes to two/more hot pairs, all these pairs are counted): $\Np$.
\end{enumerate}

\item Construct a parameter based on these gross parameters as a proxy for the randomness in the DPH. When the value of this proxy exceeds a certain cutoff, parametrized by $allowable$, then the DPH is flagged, i.e. deemed to show clustering; otherwise not.

\item If the DPH shows clustering, identify only those events in it that contribute to this flagging. Particularly, remove any lingering isolated single or double event that may have correlated with a pixel registering multiple counts, owing only to their proximity.
\end{itemize}

To optimize the values of $threshold$ and $allowable$, we resort to simulations of random DPHs, with mean count-rate of single and double events as inputs. The mean count-rate is typically $90$  events s$^{-1}$ for single and $60$   events s$^{-1}$ for double events, so in a $100$ ms timescale, they are $9$ and $6$  events  respectively. First, the number of single and double events to be chosen for a particular DPH to be simulated are drawn from Poisson distributions with the given means. Then, these many values of detx and dety are drawn from a uniform random distribution of all possible detx and dety values ($0$ to $63$). For double events, one of the neighbouring events is first chosen randomly and the other is drawn randomly from the neighbouring coordinates, taking due care of corners and edges. For the case of GRBs, the mean count-rates input into the simulation process are increased, as discussed below (see Fig. \ref{fig:robustness_with_countrate}).

For each such simulated DPH, the gross parameters $\N$, $\M$, and $\Np$ are calculated, and this is done on multiple DPHs (typically $5400$ for one full orbit) with different inputs to the parameter $threshold$. The identification of hot pairs based on $threshold$ is insensitive to the value of this parameter, as demonstrated in Fig. \ref{fig:threshold_insensitiveness}. Hence it is safe to keep it fixed at its most conservative maximum value, i.e. $0.70$, which will detect diagonally-placed neighbouring pixels each registering a count.

Next, we experiment the construction of $allowable$ based on the three gross parameters, and flag random DPHs based on the different experimental values of these parameters. It turns out that both $allowable = \M= 8$ and $allowable = \Np = 8$ flag less than $1\%$ random DPHs, but this conclusion is seen to break down in the presence of bright GRBs like GRB160802A, since the number of photons in the DPH are  $\sim 10$ times greater than the usual, resulting in random pixels getting paired and marked as hot pairs. Normalizing any of the parameters by the total number of photons does not help because extremely bright DPHstructures have total number of counts comparable to the total counts in random DPHs during GRBs, simply because the clustering illuminates its neighbourhood very brightly. Hence, we define $ allowable = \M / \N $, which normalizes for the additional $\M$ contribution from the pairs that are created due to chance co-incidence of a larger number of random events during GRBs. This simple modification fantastically  discriminates clustered DPHs from random ones,  as shown in  Fig. \ref{fig:robustness_of_allowable=3}. The reason is that, although the total number of counts in a clustered DPH is large, the clustering is spread over a few pixels, and the same pixels register many events; on the other hand, random DPHs with increased total counts, where the $\M$ is increased by co-incidental pairing of random events, have many such pairs which are themselves randomly distributed over the entire quadrant. In comparison, $allowable = \M / \Np$ does not do a better job because the small number of neighbouring pixels in a cluster tend to pair up with most of the other pixels in the cluster.

Random DPHs from GRBs and during average count-rates are examined along with DPHs that show clustering: it is seen that $allowable = \M / \N = 3$ distinctly separates clustered DPHs from random ones, whether they are during a GRB or otherwise. This is verified first visually by looking at a significant number of DPHs manually, and also demonstrated in Figs \ref{fig:robustness_of_allowable=3} and \ref{fig:robustness_with_countrate}. Examples of detected DPHstructures and DPHs with non-detections are shown in Fig. \ref{fig:DPHs_flagged_and_unflagged}.

$~$

\section*{Acknowledgements}
%I would like to sincerely thank Professor A. R. Rao, my PhD advisor in TIFR, for giving me the opportunity for doing this work, and the permission to use the L2 CZTI data for a number of sources as and when required; Professor Dipankar Bhattacharya in IUCAA for encouragement and support throughout the course of the work; NPS Mithun in Physical Research Laboratory (PRL), Ahmedabad, for his extremely helpful suggestions at crucial stages of the work, as well as long discussions clarifying doubts regarding the existing pipeline and its data files. Last but not the least, a
This publication uses data from the \A\, mission of the Indian Space Research Organization (ISRO), archived at the Indian Space Science Data Centre (ISSDC). 
The CZT Imager instrument was built by a TIFR-led consortium of institutes across India, including VSSC, ISAC, IUCAA, SAC, and PRL. The Indian Space Research Organisation funded, managed, and facilitated the project.
\bibliography{noise}   % name your BibTeX data base

\vspace{-1.5em}

%\end{theunbibliography}

\end{document}